\documentclass[cameraready]{Interspeech}

\title{Decoding while Adapting: Zero-Shot Online Speaker Adaptation via Audio-Textual Prompts for Elderly Speech Recognition}

\author[affiliation={1}]{Chengxi}{Deng}
\author[affiliation={2*}]{Xurong}{Xie}
\author[affiliation={1}]{Shujie}{Hu}
\author[affiliation={3}]{Mengzhe}{Geng}
\author[affiliation={1}]{Tianzi}{Wang}
\author[affiliation={1}]{Youjun}{Chen}
\author[affiliation={1}]{Huimeng}{Wang}
\author[affiliation={1}]{Haoning}{Xu}
\author[affiliation={1}]{Jiajun}{Deng}
\author[affiliation={1*}]{Xunying}{Liu}

\address{
    $^1$ The Chinese University of Hong Kong, Hong Kong SAR, China \\
    $^2$ Institute of Software, Chinese Academy of Sciences, China \\
    $^3$ National Research Council Canada, Canada 
}

\email{\{cxdeng, xyliu\}@se.cuhk.edu.hk, xurong@iscas.ac.cn}


\usepackage{comment}
\usepackage{multirow}
\usepackage{pifont}
\newcommand{\cmark}{\ding{51}}
\newcommand{\xmark}{\ding{55}}
\usepackage{graphicx}
\usepackage[table]{xcolor}
\usepackage{bm}
\usepackage{amsmath,amsfonts,amssymb,mathtools}
\usepackage{cite}
\usepackage[utf8]{inputenc}
\usepackage[titletoc,title]{appendix}
\usepackage{makecell}
\usepackage{tipx}
\usepackage{color}
\usepackage{multirow}
\usepackage{hhline}

\usepackage{xcolor} 

\usepackage{tabularx} 
\newcolumntype{Y}{>{\centering\arraybackslash}X} 

\begin{document}

\maketitle
\renewcommand{\thefootnote}{*}%
\footnotetext{Corresponding author.}%
\renewcommand{\thefootnote}{\arabic{footnote}}%

\begin{abstract}
This paper proposes a novel cross-utterance audio-textual prompts based speaker adaptation approach for elderly speech recognition. It enables zero-shot, real-time adaptation to unseen speakers. Speech and text embeddings are extracted from the current and a few preceding utterances, before being fused in a cross-modal manner to produce compact speaker prompts that are more consistent than i/x-vectors and ECAPA-TDNN features. Experiments on the English DementiaBank Pitt and Cantonese JCCOCC MoCA elderly speech datasets suggest that the proposed online adaptation outperforms the speaker-independent (SI) model by statistically significant word error rate (WER) or character error rate (CER) reductions of 0.61\% and 1.22\% absolute (2.99\% and 4.48\% relative). Real-time factor (RTF) speed-up ratios of up to 9.83 times are obtained over offline batch-mode adaptation.

\end{abstract}

\section{Introduction}
As populations continue to age worldwide, maintaining effective communication for the elderly plays an essential role in preserving their social engagement and quality of life. Elderly individuals confront both speech clarity issues stemming from reduced articulatory function, as well as word choice and sentence formation issues caused by weakening linguistic capabilities~\cite{becker1994natural_dbank}. The resulting reduced speech intelligibility impacts elderly individuals' daily communication and social interactions. Since existing speech foundation models are predominantly designed for normal speakers~\cite{whisper,baevski2020wav2vec,hsu2021hubert,chen2022wavlm,data2vec,QwenAudio,wavllm,llama_omini}, adapting them to accommodate elderly speech's characteristics has emerged as a crucial research challenge~\cite{hu2022exploring,ssl_shujie_taslp}.\par
Elderly speech presents multifaceted challenges to existing deep learning-based ASR technologies, including: 
\textbf{1) speaker heterogeneity}~\cite{geng2022speaker,geng2024homogeneous}: typical sources of variability, such as accent and gender, result in large speaker-level diversity among elderly speakers.
\textbf{2) data sparsity}~\cite{geng2022speaker,geng2024homogeneous,Personalized_data_aug}: mobility limitations make it difficult to collect large-scale datasets from elderly speakers.
\textbf{3) speech production and language deficiency}~\cite{deng25_interspeech}: age-related decline affects both neuromotor and neurocognitive functions in the elderly. Neuromotor deterioration leads to poor control of speech muscles, resulting in unclear articulation, reduced speaking speed, and frequent disfluencies. Meanwhile, neurocognitive decline impairs word choice and sentence formation abilities. \textbf{4) lack of longer-range, cross-utterance contextual information}~\cite{cui2025exploring}: current ASR systems for elderly speech typically require splitting the original audio into short segments before recognition. This process results in the loss of longer-range contextual information, which is essential for modeling speaker characteristics and topics.
\par
To address these challenges, various speaker adaptation methods have been proposed~\cite{geng2022speaker,geng2024homogeneous,deng25_interspeech,cuhk_elderly_zi_ye,hu2024structured,hu25d_interspeech}. Prior studies have explored the use of adapter-based~\cite{hu2024structured} and mixture-of-experts (MoE) architectures~\cite{deng25_interspeech,hu25d_interspeech} to tackle data heterogeneity challenges, including speaker-level diversity, with the goal of enabling few-shot or zero-shot capability. Meanwhile, most research~\cite{geng2022speaker,geng2024homogeneous,deng25_interspeech,hu25d_interspeech} has primarily focused on acoustic-level adaptation or separately modeling acoustic-level variations and language-level deficiencies to address speech production and language deficiency issues for the elderly.
Alternatively, utilizing cross-utterance information for ASR has proven to be an effective solution~\cite{gong2024advanced,hori2020transformer,chang2023context,duarte2024promptformer,flemotomos2025optimizing}. These methods primarily emphasize contextual relevance, such as semantic consistency. In the domain of elderly speech, existing research~\cite{cui2025exploring} relies solely on cross-utterance speech information, without explicitly performing speaker modeling to ensure speaker consistency.

\begin{figure*}[h]
    \centering    \includegraphics[width=0.8\textwidth]{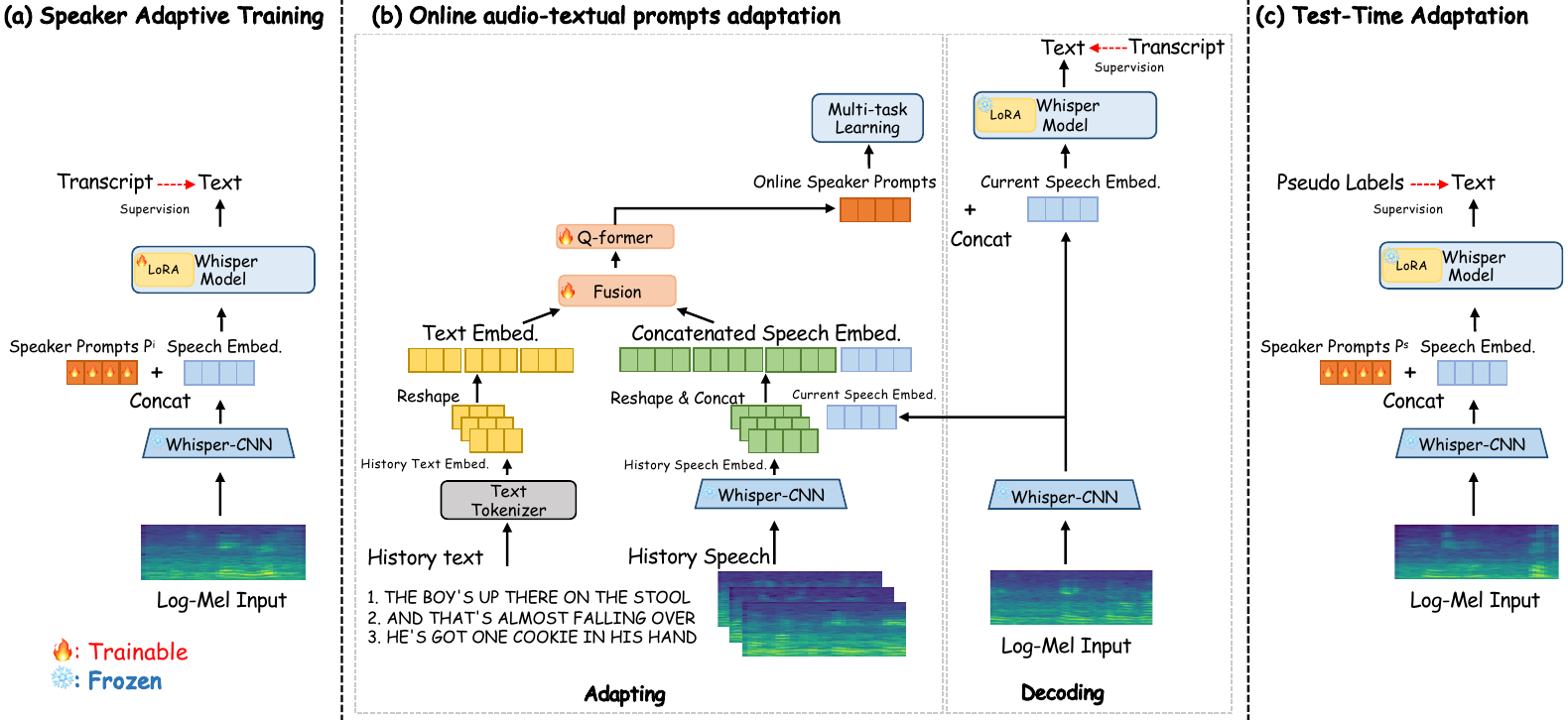}
    \caption{Examples of \textbf{(a)} the speaker adaptive training, 
    \textbf{(b)} the online audio-textual prompts adaptation,
    and \textbf{(c)} the batch-mode speaker prompts test-time adaptation.} 
\label{fig:pipeline_method}
\vspace{-0.8cm}
\end{figure*}
However, these prior studies suffer from the following limitations: {\textbf{1) Latency}}: the process of generating pseudo-labels and implementing test-time adaptation causes additional computation delays that hinder effective real-time communication for elderly users~\cite{hu2024structured}. {\textbf{2) Absence of integration between audio and textual information}}: without effective integration of contextual speech and text modalities, the estimated speaker-dependent (SD) parameters fail to simultaneously capture both the speech production and the language deficiencies, leading to inaccurate speaker modeling~\cite{geng2022speaker,geng2024homogeneous,deng25_interspeech,hu25d_interspeech,jiang24b_interspeech}.
{\textbf{3) Lack of cross-utterance textual information}}: current methods only utilize cross-utterance speech information for speaker modeling while lacking corresponding textual modality information~\cite{geng2022speaker,geng2024homogeneous,deng25_interspeech,hu25d_interspeech,jiang24b_interspeech,cui2025exploring}. The absence of textual modality leads to inaccurate speaker modeling and fails to ensure topic consistency, which is found crucial for the elderly in early detection of neurocognitive disorders~\cite{li2025detectingneurocognitivedisordersanalyses}. \par
To address these limitations, we propose a novel online speaker adaptation method that leverages history speech and textual information through dual cross-modality fusion for elderly speech recognition. This achieves three key improvements: \textbf{1)} enabling online adaptation with low latency, producing more compact speaker representations compared with i/x-vectors and ECAPA-TDNN features; \textbf{2)} integrating history speech and text contexts, jointly capturing both acoustic-level variations and language deficiencies of elderly speech for more accurate speaker modeling; \textbf{3)} compressing variable-length history speech and text information through a Q-Former based module, achieving better utilization of contextual information.

The main contributions of this work are as follows: \\
\textbf{1)} To the best of our knowledge, this is the first work employing cross-utterance audio-textual prompts for online speaker adaptation of unseen speakers in elderly speech recognition. Our contributions are threefold: \textbf{a)} Compared with batch-mode adaptation methods~\cite{hu2024structured}, our approach dynamically estimates SD parameters, achieving high performance with low latency. \textbf{b)} Compared with previous speaker adaptation methods that only focus on acoustic variability or separately modeling acoustic-level variations and language-level deficiencies~\cite{geng2022speaker,geng2024homogeneous,deng25_interspeech,hu25d_interspeech,jiang24b_interspeech}, our method effectively integrates both modalities for more accurate speaker modeling through dual cross-modality fusion. \textbf{c)} Compared with previous works that consider only a single utterance \cite{geng2022speaker,geng2024homogeneous,deng25_interspeech,hu25d_interspeech}, our approach incorporates both history speech and decoded textual 
information, achieving ``decoding while adapting" at test time. \\
\textbf{2)} Experimental results on the English DementiaBank Pitt~\cite{becker1994natural_dbank} and Cantonese JCCOCC MoCA~\cite{jccocc_datasets} elderly speech datasets suggest that the proposed method applied to the Whisper encoder outperforms the SI model by statistically significant word error rate (WER) or character error rate (CER) reductions of \textbf{0.61\%} and \textbf{1.22\%} absolute (\textbf{2.99\%} and \textbf{4.48\%} relative). Real-time factor (RTF) speed-up ratios of up to \textbf{9.83} times are obtained over offline batch-mode adaptation.
\vspace{-0.3cm}
\section{Speech Foundation Model Whisper}
\vspace{-0.2cm}
Whisper \cite{whisper} is a transformer-based encoder-decoder model trained on extensive web-scale, weakly supervised speech datasets. Leveraging the vast quantity of training data, Whisper demonstrates robust multilingual capabilities and effectively addresses a wide range of speech-related tasks. Whisper takes log-Mel spectrogram $\bm{X} \in \mathbb{R}^{D \times T}$ as input, where $D$ and $T$ respectively denote the feature dimension and input length. The convolutional block downsamples the input and the encoder subsequently transforms the downsampled input into encoder features. The decoder
then auto-regressively generates the next text tokens conditioned on the encoder features, preceding tokens and special tokens that contain the decoder prompt. 

\section{Baseline Speaker Prompt Adaptation}

\par
\noindent
\textbf{Adaptation Data Accumulation:}
Estimating speaker prompts for test speakers requires the SI system to decode the test speaker data and produce pseudo-labels for supervision. This process causes latency.
\par
\noindent
\textbf{Test-Time Adaptation:}
As illustrated in Fig. \ref{fig:pipeline_method}(c), during test-time adaptation, a set of trainable speaker prompts will be initialized for each test speaker and trained on pseudo-labels generated by the SI system. Suppose there are $S$ test speakers, the process can be formulated as follows: 
\begin{equation}
\begin{aligned}
\bm{H}_{e}^{s} &= \text{Concat}[\bm{R}^s, \bm{\text{Conv}(\bm{X})}]
\end{aligned}
\end{equation}
where $\bm{R}^s\in \mathbb{R}^{D\times L}$ represents the speaker prompts for test speaker $s$, $L$ is the length of the speaker prompts and $\bm{H}_{e}^s\in \mathbb{R}^{D\times (L+T/2)}$ denotes the hidden states after concatenating with the speaker prompts. 
\par
\noindent
\textbf{Speaker Adaptive Training:}
As shown in Fig. \ref{fig:pipeline_method}(a), for each training speaker, we initialize a set of trainable speaker prompts and perform standard speaker adaptive training (SAT) \cite{anastasakos1996compact_sat}. Considering there are $I$ training speakers, the process can be formulated as follows:  
\begin{equation}
\{\bm{\hat{R}}^i, \hat{\bm{\theta_{r}}}\} = \underset{\{\bm{R}^{i}, \bm{\theta_{r}}\}}{\arg\min} \{\mathcal{L}_{CE} (\bm{Y}^{i}|\bm{X}^i;\bm{R}^{i}, \bm{\theta}_{r})\}
\end{equation} 
where $i \in \{1,2,...,I\}$ is the index of training speakers, $\bm{R}^{i}$ is the speaker prompts of training speaker $i$. $\bm{\theta_{r}}$ is the LoRA parameters,  $\bm{X}^i$ denotes the log-Mel spectrogram input of speaker $i$, $\bm{Y}^{i}$ represents the transcript, and $\mathcal{L}_{CE}$ is the cross-entropy loss, respectively.

\begin{table*}[h]    
\caption{Performance comparison of the proposed online Audio-Textual Prompts and batch-mode prompt-based speaker adaptation with different comparable methods on DementiaBank Pitt and JCCOCC MoCA. ``Inv." and ``Par." refer to clinical investigators and elderly participants. Enc and Enc\&Dec denote applying speaker prompts to the Whisper encoder only and both encoder-decoder, respectively. Sys.8 and Sys.9 are performed only on the Whisper encoder. $^\ast$ denotes statistically significant (MAPSSWE \cite{gillick1989some}, $\alpha$ = 0.05) improvements obtained against the SI baseline ASR systems (Sys.1).}
\centering
\resizebox{\linewidth}{!}
{
    \begin{tabular}{c|c|c|c|cc|cc|c|c|c|c|c|c|c}
    \hline\hline 
    \multirow{3}{*}{Sys.} & 
    \multirow{3}{*}{Model} & 
    \multirow{3}{*}{\shortstack {Speaker\\Modeling}} & 
    \multirow{3}{*}{Online} & 
    \multicolumn{6}{c|}{DementiaBank Pitt WER(\%)} &
    \multicolumn{4}{c|}{JCCOCC MoCA CER(\%)} &\multirow{3}{*}{RTF} \\
    \cline{5-14}
    & & & & 
    \multicolumn{2}{c|}{Dev.}&\multicolumn{2}{c|}{Eval} & \multirow{2}{*}{All}&\multirow{2}{*}{\shortstack {SD\\Param.}}& \multirow{2}{*}{Dev.}& \multirow{2}{*}{Eval}&\multirow{2}{*}{All}&\multirow{2}{*}{\shortstack {SD\\Param.}} \\
    \cline{5-8}
    &&&&Par.&Inv.&Par.&Inv.&&&&&&\\
    \hline\hline
    0 & \multicolumn{3}{c|}{Conformer Transducer with Only Speech Contexts~\cite{cui2025exploring}}& 34.95&14.99&25.49&15.45&24.68&-&29.12&26.96&28.04&-&-\\
    \cline{2-4}
    1 & \multirow{9}{*}{\shortstack{Whisper\\(LoRA)}}& - & - & 28.79 & 12.76 & 20.68 & 12.65 & 20.43 & - & 28.68 & 25.76 & 27.23&-&0.24\\
    
    \cline{3-15}
    2 & & RAB \cite{hu2024structured} & \multirow{3}{*}{\xmark} & 29.54 & 13.14 & 21.27 & 12.87 & 20.99 & 0.53M & 29.35 & 26.55 & 27.94&0.53M&4.15\\
    3 & & Enc Only Prompts \cite{deng25_interspeech} & & 27.53$^\ast$&12.44&19.53$^\ast$&12.43&\textbf{19.60$^\ast$}&0.60M&27.50$^\ast$&24.32$^\ast$&\textbf{25.90$^\ast$}&0.40M&4.03\\
    4 & & Enc\&Dec Prompts \cite{deng25_interspeech} & & 27.41$^\ast$&12.05$^\ast$&19.25$^\ast$&11.76&\textbf{19.33$^\ast$}&0.75M&27.23$^\ast$&24.15$^\ast$&\textbf{25.69$^\ast$}&0.50M&4.13\\
    \cline{3-15}
    5 & & i-vector & \multirow{5}{*}{\cmark} & 29.32 & 13.13 & 21.75 & 10.99 & 20.92 & \multirow{5}{*}{-}& 39.04 & 36.16 & 37.59&\multirow{5}{*}{-}&0.27\\

    6 & & x-vector &  & 31.49 & 14.96 & 23.37 & 12.87 & 22.84 & & 29.87 & 27.43 & 28.64 &&0.27\\

    7 & & ECAPA-TDNN &  & 29.01 & 13.85 & 21.27 & 10.54 & 20.88 &  & 33.48 & 30.19 & 31.83&&0.27\\
    {\cellcolor{cyan!15}8} & & Audio-Only Prompts &  & {\cellcolor{cyan!15}28.46} & {\cellcolor{cyan!15}13.08} & {\cellcolor{cyan!15}19.73} & {\cellcolor{cyan!15}11.32} & {\cellcolor{cyan!15}20.23} & & {\cellcolor{cyan!15}28.22} & {\cellcolor{cyan!15}25.32} & {\cellcolor{cyan!15}26.76}&&{\cellcolor{cyan!15}0.40}\\
    {\cellcolor{orange!20} 9} & & Audio-Textual Prompts &  & {\cellcolor{orange!20}28.05} & {\cellcolor{orange!20}12.51} & {\cellcolor{orange!20}19.65$^\ast$} & {\cellcolor{orange!20}11.21} & {\cellcolor{orange!20}\textbf{19.82$^\ast$}}& & {\cellcolor{orange!20}27.26} $^\ast$& {\cellcolor{orange!20}24.78 $^\ast$}& {\cellcolor{orange!20}\textbf{26.01$^\ast$}}&&{\cellcolor{orange!20}0.42}\\

    \hline\hline
\end{tabular}
}
\label{tab:table_online}
\end{table*}

\section{Online Speaker Adaptation using Audio-Textual Prompts}
\par
\begin{figure}[h]
    \centering
    \includegraphics[width=0.45\textwidth]{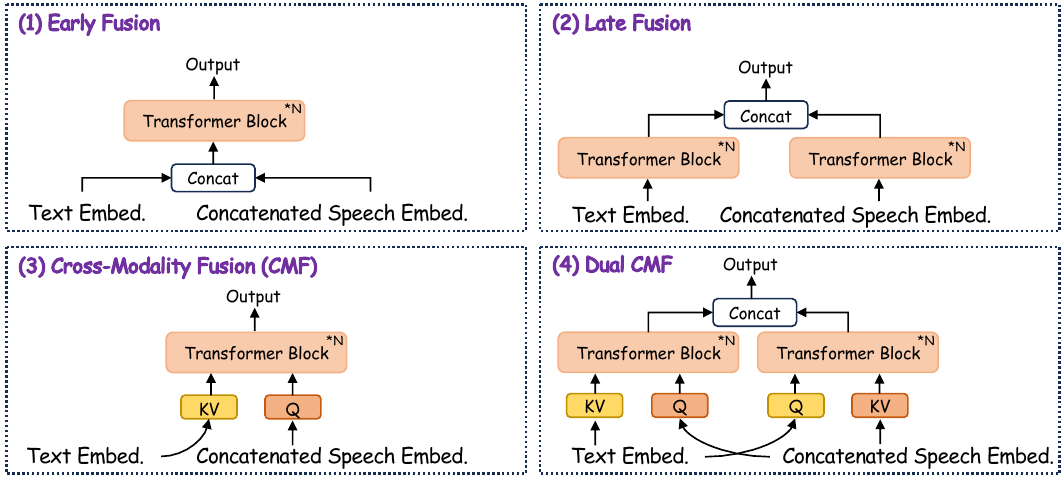}
    \caption{The proposed different fusion strategies for audio and textual information: (1) Early Fusion; (2) Late Fusion; (3) Cross-Modality Fusion (CMF); and (4) Dual Cross-Modality Fusion (Dual CMF).}
\label{fig:fusion_method}
\end{figure}

\noindent
\textbf{Audio-Textual Prompts with Cross-Utterance Contexts:}
The online method requires a two-step training process as shown in Fig. \ref{fig:pipeline_method}(a)\&(b). As illustrated in Fig. \ref{fig:pipeline_method}(b), the history speech is encoded using the same method as the current speech and the history speech features are concatenated with the current speech features along the temporal dimension\footnote{The optimal concatenation position for the online speaker prompts is after the Whisper encoder for the DementiaBank Pitt dataset and after the Whisper CNN for the JCCOCC MoCA dataset.}. History text is encoded using the Whisper tokenizer, with ground-truth transcripts for training and decoded text for inference. As shown in Fig. \ref{fig:fusion_method}, four types of fusion methods between speech and text embeddings are investigated: 
\textbf{(1) Early Fusion}, which directly concatenates history text and speech embeddings along the temporal dimension and processes them through three transformer blocks to generate fused representations; \textbf{(2) Late Fusion}, where text and speech embeddings are processed separately through three individual transformer blocks before being concatenated; \textbf{(3) Cross-Modality Fusion (CMF)}, which uses history speech embeddings as Query vectors and history text embeddings as Key/Value vectors in three cross-attention layers; and \textbf{(4) Dual Cross-Modality Fusion (Dual CMF)}, which employs three dual cross-attention layers, each taking one modality as the Query vectors and the other modality as the Key/Value vectors. All four approaches utilize the same Q-Former architecture \cite{li2023blip} to compress the fused history context features. The utilized Q-Former consists of 8 transformer blocks, each containing self-attention and cross-attention layers followed by feed-forward layers with residual connections \cite{attention}. Specifically, we initialize a set of learnable query vectors and utilize self-attention to model inter-query relationships, then perform cross-attention with the variable-length fused history contexts features to extract contextual information. The Q-Former enables adaptive extraction from the fused history audio-textual features, capturing both acoustic-level variations and language deficiencies when performing speaker modeling. For the audio-only prompts systems, no fusion strategies are applied and we directly use the Q-Former to extract speaker prompts from the speech. \par

\par
\noindent
\textbf{Multi-Task Learning:}
We design three loss functions for training. To ensure performance on the ASR task, \textbf{1)} cross-entropy loss $\mathcal{L}_{ASR}$ is applied, which is the same as the original training loss \cite{whisper}.
To maintain speaker identity consistency, \textbf{2)} an auxiliary speaker classification task with a cross-entropy loss $\mathcal{L}_{Spk}$ is used. The speaker classification module first performs temporal averaging, then processes the features through four linear layers with dimensions of 1000, 1000, 1000, and the number of target speaker classes\footnote{688 for the DementiaBank Pitt dataset and 369 for the JCCOCC MoCA dataset.}, with each linear layer followed by dropout and ReLU activation, and finally applies a softmax layer for speaker classification. To ensure consistency between the online speaker prompts and speaker prompts directly estimated by offline SAT, \textbf{3)} a mean squared error (MSE) loss $\mathcal{L}_{MSE}$ is applied to enforce alignment between them. The overall loss function is given as:
$\mathcal{L}_{All} = \mathcal{L}_{ASR}+\alpha\mathcal{L}_{Spk}+\beta\mathcal{L}_{MSE}\footnote{$\alpha$ and $\beta$ are empirically set to 2 and 0.2, respectively.}$. 
\par
\noindent
\textbf{Decoding While Adapting:}
We achieve decoding while adapting for inference. During inference, we utilize the already decoded text obtained through greedy search as history textual information. The history text is then fused with the corresponding speech to generate online speaker prompts on the Whisper encoder side, which are subsequently utilized for decoding the current utterance. Unlike batch-mode adaptation which requires pre-decoded text as pseudo-labels and performs test-time adaptation (Fig. \ref{fig:pipeline_method}(c)) to generate speaker prompts, our method dynamically leverages audio-textual information during decoding to generate online speaker prompts. This approach achieves faster processing than batch-mode adaptation.

\section{Experiments}
\par
\noindent
\textbf{Task Description:}
The English \textbf{DementiaBank Pitt} \cite{becker1994natural_dbank} corpus is the most widely used publicly available elderly speech corpus for Alzheimer’s Disease (AD) diagnosis. The dataset consists of 33 hours of audio recordings collected from 292 interviews involving elderly participants and clinical researchers conducting Alzheimer's disease assessments. The training set includes 688 speakers, while the development and evaluation sets contain 119 speakers and 95 speakers, respectively. After applying silence stripping and data augmentation techniques \cite{cuhk_elderly_zi_ye}, the training set expands to 58.9 hours, with the development and evaluation sets including 2.5 hours and 0.6 hours of audio, respectively. The Cantonese \textbf{JCCOCC MoCA} \cite{jccocc_datasets} corpus comprises 256 cognitive impairment assessment interviews between elderly participants and clinical investigators. The training set includes 369 speakers, while the development and evaluation sets each consists of recordings from two different groups of 49 elderly speakers. \textbf{No elderly speakers in the training set overlap with those in the development or evaluation sets for both datasets.} \par

\par
\noindent
\textbf{Experimental Setup:}
We adopt Whisper-medium\footnote{https://huggingface.co/openai/whisper-medium} for its multilingual capabilities and robust generalization performance. We directly perform inference using the original Whisper-medium, obtaining overall WER (CER) of \textbf{28.01\%} and \textbf{93.57\%} on DementiaBank Pitt and JCCOCC MoCA, respectively. We apply LoRA \cite{lora} to the “query”, “key”, “value”, and “att.out” of the attention module, with the LoRA rank set to 8. \textbf{The LoRA-based fine-tuning on Whisper achieves state-of-the-art baseline performance on both datasets, outperforming other foundation models~\cite{ssl_shujie_taslp}.} Further online speaker adaptation methods are conducted based on this state-of-the-art baseline system. 

\begin{table}[t] 
\caption{Performance of online adaptation. ``History Info.'' refers to ``History Information''. ``utt.'' denotes ``utterance".}
    \vspace{-0.3cm}
    \centering
    \renewcommand{\arraystretch}{1.15} 
    \resizebox{\linewidth}{!}
    {
    \begin{tabular}{c|c|c|c|c|cc|cc|c}
    \hline\hline 
    \multirow{3}{*}{Sys.} & 
    \multicolumn{3}{c|}{History Info.} & 
    \multirow{3}{*}{\shortstack {Modality\\Fusion}} & 
    \multicolumn{5}{c}{DementiaBank Pitt WER(\%)} \\
    \cline{2-4}
    \cline{6-10}
    &\multirow{2}{*}{\shortstack{Prompt\\Length}}&
    \multirow{2}{*}{\shortstack{Speech}}
    &
    \multirow{2}{*}{Text}
    &&
    \multicolumn{2}{c|}{Dev.} & \multicolumn{2}{c|}{Eval.} & \multirow{2}{*}{All} \\
    \cline{6-9}
    &&&&&
    Par. &Inv. &Par. &Inv. \\
    \hline\hline

1&1&\multirow{6}{*}{0 utt.}&\multirow{9}{*}{0 utt.}&\multirow{9}{*}{-}&
29.03&12.42&20.55&12.76&20.37  \\
2&2&&&&
28.39&12.59&20.79&14.21&20.26 \\
3&4&&&&
28.61&12.52&20.16&12.10&20.16 \\
4&8&&&&
28.64&12.57&20.09&10.99&20.14  \\
5&16&&&&
28.48&12.69&20.49&10.99&20.19 \\
6&32&&&&
28.64&12.84&20.89&12.65&20.43 \\
\cline{1-3}
7&\multirow{9}{*}{8}& 1 utt.&&&
28.45&12.82&19.97&11.21&20.15 \\
{\cellcolor{cyan!15}8}&& 3 utt.&&&
    {\cellcolor{cyan!15}28.46}&{\cellcolor{cyan!15}13.08}&{\cellcolor{cyan!15}19.73}&{\cellcolor{cyan!15}11.32}&{\cellcolor{cyan!15}20.23} \\

9&& 5 utt.&&&
28.39&13.01&19.67&11.65&20.15 \\
\cline{3-10}
10&& 1 utt.& 1 utt.&\multirow{3}{*}{Dual CMF}&
28.12&12.73&19.78&11.32&19.96 \\
{\cellcolor{orange!20} 11}&& 3 utt.& 3 utt.&&
{\cellcolor{orange!20}28.05}&{\cellcolor{orange!20}12.51}&{\cellcolor{orange!20}19.65}&{\cellcolor{orange!20}11.21}&{\cellcolor{orange!20}19.82} \\

12&& 5 utt.& 5 utt.&&
28.03&12.84&19.76&11.43&19.97 \\
\cline{3-5}
13&&\multirow{3}{*}{3 utt.}&\multirow{3}{*}{3 utt.}&Early Fusion&
28.41&12.78&19.97&11.21&20.12  \\
14&&&&Late Fusion&
28.42&12.82&19.86&11.21&20.13  \\
15&&&&CMF&
28.49&12.61&19.82&11.10&20.06  \\
\cline{5-10}
\hline\hline
\end{tabular}}
\label{tab:table_ablation}
\end{table}
\vspace{-0.2cm}
\begin{figure}[h]
    \centering        
    \includegraphics[width=0.3\textwidth]{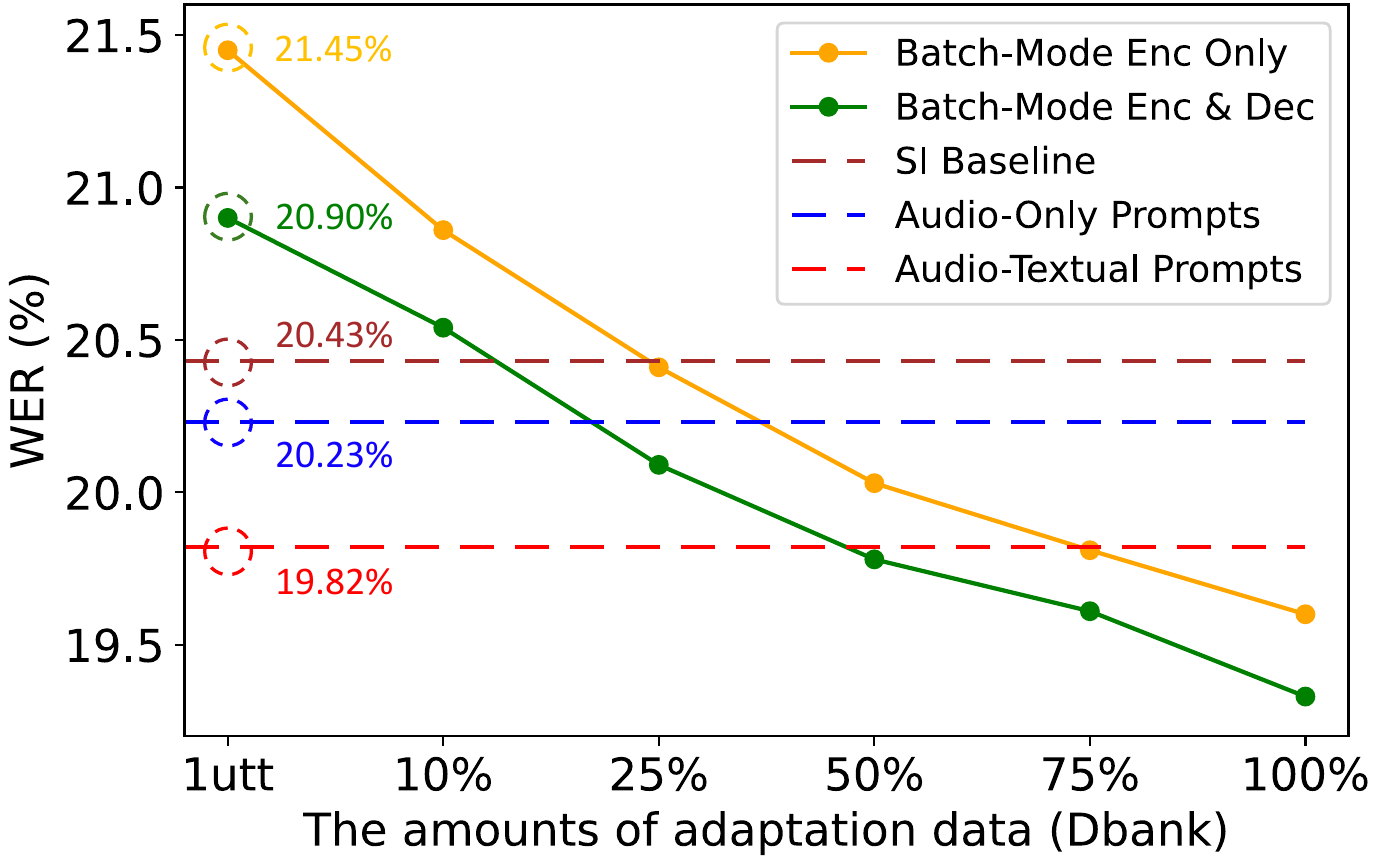}
    \caption{WER on adapted Whisper systems with varying amounts of speaker adaptation data on the DementiaBank (Dev+Eval) data.}
\vspace{-0.4cm}
\label{fig:data_amount}
\end{figure}
\par
\begin{figure}[h]
    \centering        \includegraphics[width=0.45\textwidth]{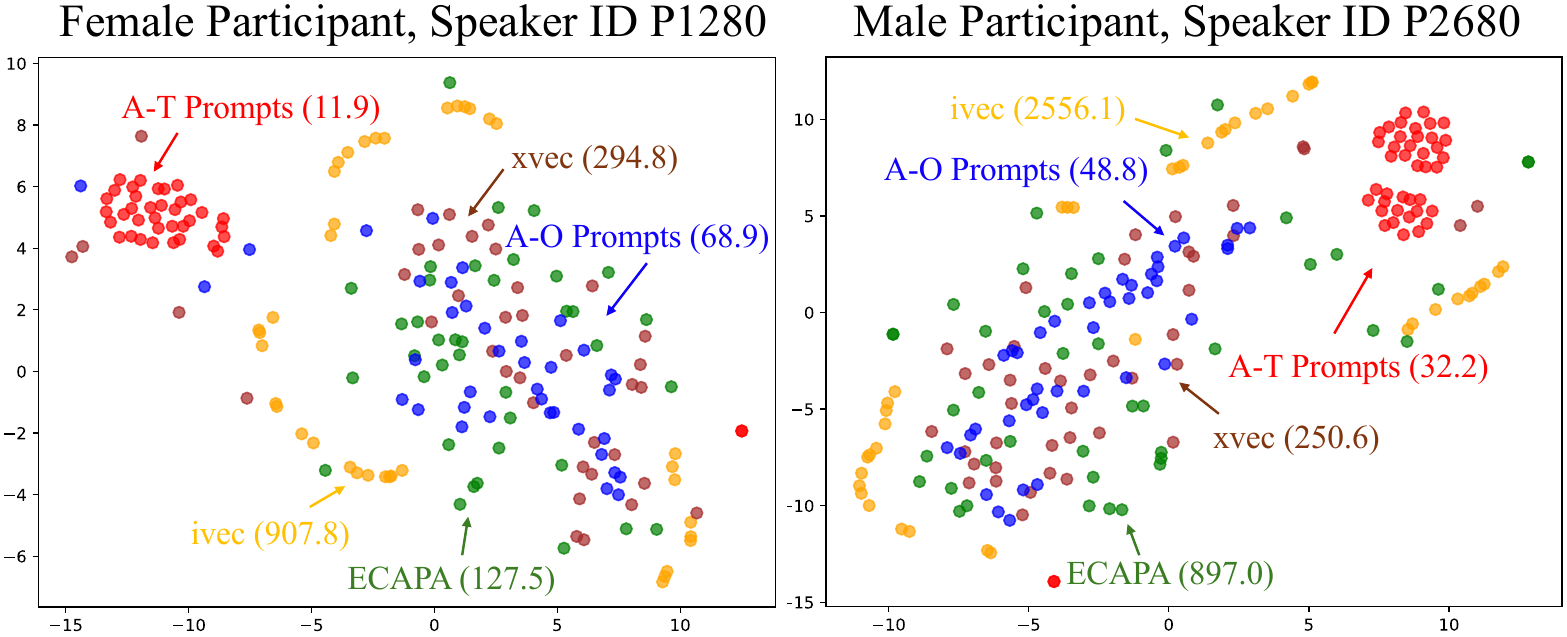}
    \vspace{-0.2cm}
    \caption{T-SNE visualization of the Audio-Textual Prompts (A-T Prompts), Audio-Only Prompts (A-O Prompts), i/x-vectors (i/xvec) and ECAPA-TDNN (ECAPA) features (Table \ref{tab:table_online}, Sys.5-9) for 2 speakers in DementiaBank Eval set. Each bracket shows determinants of the covariance matrix.}

\label{fig:tsne_result}
\end{figure}

\noindent
\textbf{Zero-Shot Performance:}
Since there is no overlap between training and test speakers, the proposed method operates under zero-shot conditions. From the experimental results, several trends can be found:\\
\textbf{1)} Compared with the batch-mode encoder only prompts and the encoder\&decoder prompts adaptation, our online audio-only prompts and audio-textual prompts adaptation are invariant against speaker data quantity as shown in Fig. \ref{fig:data_amount} (blue dashed line vs. yellow solid line and red dashed line vs. green solid line). Our audio-textual prompts adaptation achieves comparable performance to batch-mode encoder\&decoder prompts adaptation, while delivering \textbf{9.83} times RTF speedup in inference (Table \ref{tab:table_online}, Sys.9 vs. Sys.4). Specifically, the proposed audio-textual prompts outperform the baseline (Table \ref{tab:table_online}, Sys.9 vs. 1) with statistically significant WER and CER reductions of \textbf{0.61\%} and \textbf{1.22\%} absolute (\textbf{2.99\%} and \textbf{4.48\%} relative) on DementiaBank Pitt and JCCOCC MoCA, respectively. We also show the performance of conformer transducer using only speech contexts (Table \ref{tab:table_online}, Sys.0).\\
\textbf{2)} Compared with i/x-vectors and ECAPA-TDNN adaptation, our online audio-only prompts and audio-textual prompts adaptation 
achieve better performance (Table \ref{tab:table_online}, Sys.8-9 vs. Sys.5-7). These gains are consistent with the T-SNE visualization in Fig. \ref{fig:tsne_result}, where the speaker representations produced by audio-only prompts and audio-textual prompts are \textbf{more consistent} than those obtained from i/x-vectors and ECAPA-TDNN. \\
\textbf{3)} Compared with audio-only prompts adaptation, the proposed audio-textual prompts adaptation achieves superior performance across different history contexts lengths (Table \ref{tab:table_ablation}, Sys.10-12 vs. Sys.7-9) and this performance advantage remains consistent across both datasets (Table \ref{tab:table_online}, Sys.9 vs. Sys.8). These results indicate that incorporating the textual modality leads to more effective speaker modeling by providing additional contextual information that complements the acoustic features, and these gains are also consistent with the T-SNE visualization in Fig. \ref{fig:tsne_result}, where speaker representations produced by audio-textual prompts adaptation are more consistent than audio-only prompts adaptation.
\vspace{-0.3cm}

\vspace{-0.6cm}
\par
\noindent
\textbf{Ablation Studies:}
As shown in Table \ref{tab:table_ablation}, different configurations of online methods are investigated on DementiaBank Pitt, including: \textbf{1) Prompt Length}. The optimal results are obtained using a prompt length of 8 for online speaker prompts generation using only current speech (Sys.4 vs. Sys. 1-3, 5-6). \textbf{2) History Utterance Count}. The optimal configuration uses 3 pairs of history speech and decoded text (Sys.11 vs. 10, 12). \textbf{3) Fusion Strategy}. The proposed \textbf{dual cross-modality fusion method (Dual CMF)} demonstrates superior performance compared with the other three fusion approaches, suggesting that this fusion method can more effectively integrate history speech and text information (Sys.11 vs. Sys.13-15). The above optimal settings are employed for the experiments reported in Table \ref{tab:table_online} on both datasets.
\vspace{-0.4cm}
\section{Conclusion}
This paper proposes a novel online speaker
adaptation method that leverages history speech and textual information for elderly speech recognition. Cross-utterance contextual information is fused through dual cross-modality fusion and effectively compressed using Q-Former for speaker modeling. Experimental results on DementiaBank Pitt and JCCOCC MoCA elderly speech datasets suggest that our method outperforms the SI baseline by WER and CER reductions of 0.61\% and 1.22\% absolute (2.99\% and 4.48\% relative). Compared to offline batch-mode adaptation, the proposed method achieves RTF speed-up ratios of up to 9.83 times.

\section{Acknowledgements}
This research is supported by Hong Kong RGC GRF grant No. 
14200021, 14200324, 
Basic Research Project of Institute of Software, Chinese Academy of Sciences ISCAS-JCMS-202306, and Youth Innovation Promotion Association CAS Grant 2023119.

\section{Generative AI Use Disclosure}
Generative AI tools were used only for language editing and improving readability during the preparation of this manuscript. These tools were not used to generate core scientific ideas, experimental data, or technical contributions. All authors have thoroughly reviewed and approved the final manuscript and assume full responsibility for the integrity of its entire content.

\bibliographystyle{IEEEtran}
\bibliography{mybib}

\end{document}